# The Caloric Curve for Mononuclear Configurations

L.G. Sobotka[1,2], R.J. Charity[1], J. Tõke[3] and W. U. Schröder[3]

*Departments of Chemistry[1] and Physics[2], Washington University, St. Louis, MO. 63130 and*
*[3]Department of Chemistry, University of Rochester, Rochester, NY 14627}*
(Dated: November 13, 2018)

The caloric curve for mononuclear configurations is studied with a schematic model. We investigate the dependence of the entropy on the density and effective mass profiles. A plateau in the caloric curve is a direct result of decreasing density and the destruction of correlations with increasing excitation. The mononuclear regime is metastable with respect to binary fission at low excitation energy and unstable with respect to multifragmentation at high excitation. The statistical framework presented here is suitable to treat scenarios where experimental conditions are set to favor population of highly excited mononuclear states.

Heavy compound nuclei (CN) are metastable objects; local mononuclear entropy maxima separated from the more stable dinuclear states by transition states of significantly lower entropy. Standard statistical-model treatments of CN decay are predicated on a time-scale separation between the CN formation time, and the time scales for simple (mostly single-particle) decay modes and the massively collective decay processes we call fission. The former are usually treated using the prescriptions offered by Weisskopf[1] or Hauser and Feshbach[2] while the latter are usually treated by a transition-state formalism initially developed for chemical reactions by Eyring[3]. The distinction between decay modes can be bridged in concept[4] and practice[5] however the massively collective decay channels can be retarded by transient delays[6].

It has been known for almost 4 decades that below approximately 1/3 of the saturation density $\rho_o$, $\alpha$-matter has a lower free energy than uniform nuclear matter[7][8]. However mononuclear configurations at reduced density can be reasonable subjects for statistical decay treatments as long as they are either metastable or protected from massively collective decays modes by transient delays. Specifically with increasing excitation energy, an equilibrium (i.e. local maximal entropy) mononuclear density profile can be reached, the decay of which can be treated with only minor modifications to the well known formalisms, as long as a time-scale separation exists between formation, equilibration and all conceivable decay modes.

This work does not deal with the important issue of time-scale separation[9], only with the gross statistical properties of mononuclei at high excitation energy. In particular, we show that the relaxation of the density profile of mononuclei, in pursuit of maximal entropy, causes the Caloric curve, $T(\varepsilon)$ (where $T$ is the statistical nuclear temperature and $\varepsilon$ is the excitation per nucleon) to flatten out and exhibit a quasi-plateau. We believe this explains the nature of the Caloric curve first studied by Wada et al.[10] and later by Pochodzalla et al.[11], and for which systematics have recently been analyzed in detail by Natowitz et al. [12].

The approximate saturation of the statistical temperature is primarily due to density reduction, but is also influenced by the evolution of the effective mass of nucleons in the nuclear medium. The first effect is just the sequestration of energy in the potential energy of nuclear expansion. The energy spent on expansion reduces the thermal part of the total excitation in much the same way as the collective rotational energy does in the case of high angular momentum CN.

The ratio of the effective mass to the bare nucleon mass $m^*/m$ differs from 1 due to the finite range of the nuclear force and the time non-locality of the interaction. The former effect, which is responsible for making the optical model potential energy dependent, reduces $m^*/m$ by a density dependent factor $m_k(\rho)$ which must return to 1 at low density. The time nonlocality can be thought of as the coupling of low-lying surface modes to single-particle degrees of freedom[13][14]. This collective effect brings strength down from high energy, increasing the many-body density of states at low excitation energy. The effective mass factor, $m_\omega(\rho',T)$, accounting for this relocation of strength, while greater than 1 at low energy and localized on the surface of the quantum drop, must return to 1 in the limit of high excitation as well as low density gradient.

We confine our analysis to a one parameter description of expansion and literature descriptions of how the effective-mass terms evolve with density and excitation. Our approach combines the physically transparent picture of maximal-entropy mononuclear configurations found in the recent work by Tõke et al. [15], with effective-mass logic similar to that found in the works of Natowitz, Shlomo and collaborators[16].

A quantum drop of degenerate Fermi liquid has an entropy given by[17],

$$S_M = 2\sqrt{aU} = 2\sqrt{a(E_T^* - E_C)} = 2\sqrt{aA(\varepsilon - \varepsilon_C)}, \quad (1)$$

where $a$ is the level-density parameter and the thermal, total and compressional energies are $U$, $E_T^*$, and $E_C$, respectively. With total particle number $A$, $\varepsilon$ and $\varepsilon_C$ are the total and compressional energies per particle. In the local density approximation (LDA)[18], the level density depends on the nuclear profile, the local Fermi momentum $k$ and the effective mass[19][20],

$$a = \frac{\pi^2}{4} \sum_\tau \int \frac{\rho_\tau(r)}{[\hbar^2 k_\tau^2(r)/2m^*]} d\mathbf{r}. \qquad (2)$$

The density profiles $\rho_\tau(r)$ of the two isospin partners (with index $\tau$) are taken to be the same functional form, scaled in proportion to the number of nucleons. The native ($\epsilon = 0$ MeV) radial profiles are of the "standard" type with a Gaussian derivative,

$$\rho_n(r) = \frac{\rho_o}{2}(1 - \text{erf}(\frac{r - R_o}{\sqrt{2}b})), \qquad (3)$$

with effective sharp radius $R_o = r_o A^{1/3}$ ($r_o = 1.16$ fm) and surface width $b = 1.0$ fm. The expansion is limited to the one-dimensional self-similar family, i.e. $\rho(r,c) = c^3\rho_n(c*r)$. The expansion parameter $c$ is found by maximizing the mononuclear entropy,

$$\frac{\partial S_M(E_T^*, \rho(c))}{\partial c}]_{E_T^*} = 0. \qquad (4)$$

The collective energy involved in expansion is taken as the simple parabolic form, involving only the central density, suggested by Friedman[21], $E_c = A*\varepsilon_b(1-\frac{\rho(0)}{\rho_o})^2$. We have used $\varepsilon_b = 8$ and 6 MeV (red and green curves in figures 2-4) in the present calculations[22].

Execution of eq. 4, not only finds the metastable mononuclear expansion but also insures that the surface pressure is zero. This procedure is therefore logically different from the physically unreal, but true equilibrium condition found by placing a drop in a box and having a surrounding vapor supply a pressure.

Following Prakash et al. [20], and De et al. [24] we choose a phenomenological form for the effective mass that captures the underlying science,

$$\frac{m^*}{m} = (m_k)[m_\omega] = (1 - \alpha\frac{\rho(r)}{\rho_o})[1 - \beta(T)\frac{\rho'(r)}{\rho_o}], \qquad (5)$$

with

$$\alpha = 0.3, \ \beta(T) = 0.4A^{1/3}\exp[-(TA^{1/3}/21)^2]. \qquad (6)$$

(The $T$ dependence in $\beta(T)$ requires performing an iteration of the outlined procedure. This modification is straightforward and will not be described.) The effective-mass factor is suppressed in the bulk, peaks at the surface[25] and degrades to 1 with decreasing density and increasing thermal energy. These two many-body effects, to a large extent, offset one another in near ground-state nuclei, yielding $a \approx A/8.6$ for unexpanded $^{197}$Au, the nucleus considered here. However the destruction of the cooperativity encoded in these two effective-mass terms does not occur on identical energy scales. While the detailed density and excitation energy dependence of these terms is unknown, the present work shows how the gross effects captured by these terms couple with expansion to dictate the form of the Caloric curve.

The excess entropies, $S_M^E = S_M - S_M(c = 1)$, are shown in Fig. 1. The maximum entropy determines the equilibrium expansion parameter ($c$) as a function of $\varepsilon$.

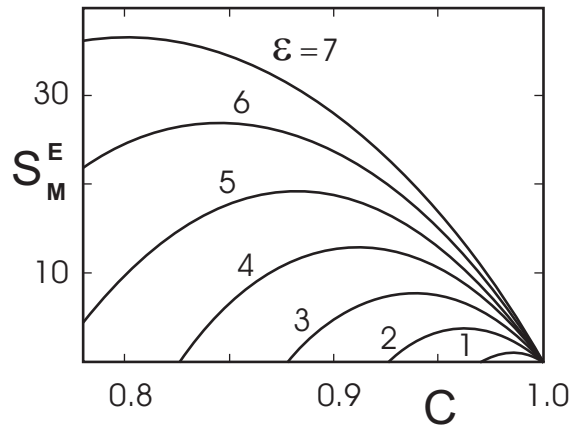

Fig. 1: Representative calculation of excess entropy as a function of expansion: $S_M^E(c) = S_M - S_M(c = 1)$.

The reduction of central density with excitation $\rho_c(\varepsilon)$ is shown in Fig. 2 without $[m^* = 1]$ (black) and with $[m^*/m = m_k(\rho)m_\omega(\rho', T)]$ (red: $\varepsilon_b = 8$ MeV and green: $\varepsilon_b = 6$ MeV) effective mass considerations. (Consideration of $m_k(\rho)$ alone exhibits a reduction in the central density similar to that with $m_k(\rho)m_\omega(\rho', T)$, while consideration of $m_k(\rho)m_\omega(\rho')$ leads to approximately the same $\rho_c(\varepsilon)$ as with $m^*/m = 1$.) Without the effective-mass terms, the $\rho_c(\varepsilon)$ dependence is almost identical with the (extended) finite-temperature Hartree-Fock calculation reported in [16]. As one should expect, the decrease in density is more substantial with the smaller compressibility. The central densities implied from the maximal-entropy procedure used here with $m^*/m = m_k(\rho)m_\omega(\rho', T)$ and $\varepsilon_b = 6 \ MeV$ are similar to those extracted from caloric curve data in [16] (diamonds). On the other hand, $\rho_c(\varepsilon)$ does not drop quickly, over a narrow a range of energies, as is suggested by the analysis of Coulomb barriers by Bracken et al. [27] (circles).

Figure 3 displays $S_M$ for the native density profile (dotted), as well as $S_M$ and $S_M^E$ for $m^*/m = 1$ (black), $m^*/m = m_k(\rho)m_\omega(\rho', T)$ for $\varepsilon_b = 8$ MeV (red), and $\varepsilon_b = 6$ MeV (green). The excess entropy allowed by relaxing the density profile (black dashed) is evident as is a reduced rate of entropy growth at for $\varepsilon < 4$ MeV when the effective mass is modelled by equations 5 and 6. This "reduction", is due to decreasing $m^*/m$ in the surface region with increasing $\varepsilon$.

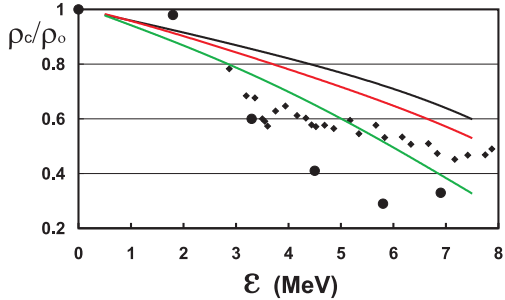

Fig. 2: Central densities for $^{197}$Au with $[m^* = 1, \varepsilon_b = 8$ MeV] (black), $[m^* = m_k(\rho)m_\omega(\rho', T), \varepsilon_b = 8$ MeV] (red), and $[m^* = m_k(\rho)m_\omega(\rho', T), \varepsilon_b = 6$ MeV] (green). Also shown are the densities extracted from apparent level density parameters (diamonds) [26] and from Coulomb barriers (circles) [27].

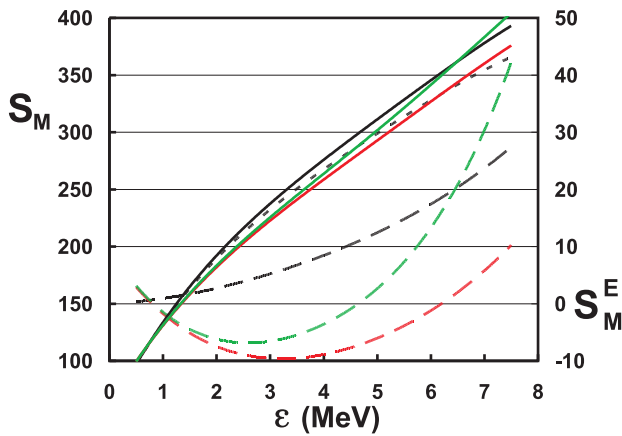

Fig. 3: $S_M$ for a native density profile (dotted), and $S_M$ (solid) and $S_M^E$ (dashed) for the cases shown in Fig. 2.

The statistical temperatures, found by differentiation,

$$T = A/[\frac{\partial S}{\partial \varepsilon}]_v, \tag{7}$$

are shown in Fig. 4. (The nuclear volume is not a thermodynamic control variable, rather a volume sufficiently large to make the pressure zero is. This volume is unaffected by the changes in the nuclear volume. From another perspective, as the external pressure is zero, no work is done by the nuclear expansion.) As shown in [15] and inferred in [16], the relaxation of the density profile, substantially flattens the temperature rise with $\varepsilon$ (compare black dashed and solid curves). The inclusion of the full effective mass form (function of $\rho$, $\rho'$, and $T$) increases

$T$ for $\varepsilon < 4$ MeV and decreases it for $\varepsilon > 5$ MeV (compare solid black to colored curves), changes that give the appearance of a plateau. Decreasing the compressional energy constant (red to green), reduces the value of the temperature of the pseudo-plateau.

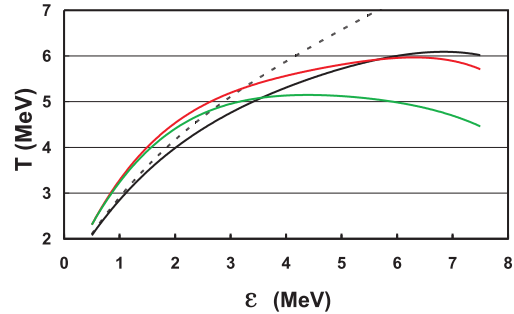

Fig. 4: $T$ for the same cases as shown in the previous two figures.

Thermodynamic stability of a single phase requires the total $S_T(E)$ have a positive first derivative (positive temperature) as well as being concave (implying a positive heat capacity $C_v$ as $(\frac{\partial^2 S}{\partial E^2})_v = -1/T^2 C_v$). However the same cannot be said of the constrained mononuclear $S_M(E)$. Comparison of $S_M(E)$ to $S_{MF}$, the entropies for multifragmentation, is complicated by the prescription for the free volume needed for the latter[28][29]. Nevertheless one can compare $S_M(m^*/m = 1)$, at extracted temperatures, to $S_{MF}$ as a function of volume. Taking the latter from the ideal phase space model of Das Gupta and collaborators [30] indicates: 1) The volume capturing 99.75% of the mononuclear matter is much smaller than reasonable volumes for multifragmentation. 2) $S_{MF}$ will exceed $S_M$ somewhere above $\varepsilon > 3 \ MeV$ if a freezeout volume of 3 times the unexpanded volume is used. This crossing implies that the mononucleus becomes unstable with respect to the multifragmented state trapped in a hypothetical freezeout volume. As the latter occupies a much larger volume, the multiparticle phase space regions are divorced indicating that transient times, as well as pure statistical rates, are needed in a decay treatment.

An irreversible phase change (finite free energy change) from an initially populated metastable or kinetically trapped mononucleus to the higher entropy multifragmented state would proceed with evolution of the kinetic energy and thus event averages, sampling a time distribution, would yield fluctuations in excess of those expected for an equilibrium system. We therefore suggest that excessive fluctuations of the kinetic energy [31], result from an irreversible transition from a lower entropy (larger free energy in a canonical treatment) but kinetically trapped, mononuclear phase to the multifragmented phase. This is equivalent to arguing that the ensemble (the event sample) is non-ergodic. However this transition is not required to explain a plateau in a caloric curve, nor are large fragment multiplicities or significant final-state correlations.

As concerns fragment multiplicity, the characteristic (evaporative) decay time $\tau_{ev} = \frac{\hbar}{\Gamma}$, with the decay width calculated using realistic, but simple forms for the neutron decay width $\Gamma_n$ [32], is of the order of 10-15 fm/c across the plateau. In a standard (binary transition-state) statistical model, many fragments would be evaporated as the CN retraces the Caloric curve going to the left, contracting as the sequence continues. As the temperature does not change significantly across the plateau, the issue of sequence is irrelevant.

The relevance of such a statistical model depends on there being a time-scale separation between mononucleus formation $\tau_{for}$ and the decay times associated with mononuclear $\tau_{ev}$ and multifragment $\tau_{mf}$ final states. It is unlikely that $\tau_{for} < \tau_{ev}$ in heavy-ion reactions, but this condition might be met for light ion, $\pi$ and $\tilde{p}$ induced reactions. The time scale relevant for $\tau_{mf}$ is the maximum of a purely statistical time, related to the entropy of the transition state $S_{MF}$, and the delay time associated with the amplification of density fluctuations. Work on the difficult issue of the spinodal decomposition time scale[9] suggests that $\tau_{ev} < \tau_{mf}$ in the plateau. A clear analogy between the transient delay for fission[6] and the amplification time for density fluctuations leading to spinodal decomposition can be made.

The present work indicates that a purely statistical treatment of the decay of a highly excited nucleus (presuming formation of an equilibrated mononucleus) would be one which evaporates from metastable (local maximal entropy) expanded systems and has a decay branch with transient delay for spinodal decomposition. The significance of the delay time is that even if $S_{MF} > S_M$ the decay might still be determined by the mononuclear phase space. We note that the logic for treatment of the evaporative-decay channels outlined above is substantially different from that offered in the "Expanding Emitting Source" model proposed by Friedman[21]. In EES, the expansion is dynamically produced rather than the result of maximizing the entropy.

This work makes use of several simplistic assumptions. The LDA is known to yield inaccurate level density results[18] and the self-similar expansion is a restriction that limits the entropy. The expression for the entropy is also simplistic in that it ignores the continuum and all detailed quantum structure. Nevertheless, it seems that a near plateau of the Caloric curve is an unavoidable consequence of an expanding metastable Fermion system (with a finite-range force) and the resultant evolution (destruction) of collectivity with expansion and excitation.

LGS would like to acknowledge fruitful discussions with Professors W. Dickhoff, and R. Lovett. This work was supported by the U.S. Department of Energy, Division of Nuclear Physics under grant DE-FG02-87ER-40316 and DE-FG02-88ER-40414, for Washington University and University of Rochester, respectively.